\documentclass[twocolumn,prl,aps,superscriptaddress,showpacs,amsmath,amssymb,floatfix]{revtex4-1}
\usepackage{color}
\usepackage{xcolor}
\usepackage{mathrsfs}
\usepackage{amsmath}
\usepackage{graphicx}
\usepackage{dcolumn}
\usepackage{bm}
\usepackage{times}
\usepackage{amssymb}
\usepackage{float}
\usepackage{array}
\usepackage{tikz}

\usepackage{ulem}

\begin{document}

\title{Fundamental intrinsic lifetimes in semiconductor self-assembled quantum dots}
\author{Wen Xiong}
\affiliation{Department of Physics, Chongqing University, Chongqing 401331, China}
\author{Jun-Wei Luo}
\thanks{Email: jwluo@semi.ac.cn}
\affiliation{State Key Laboratory of Superlattices and Microstructures, Institute of Semiconductors, Chinese Academy of Sciences,  Beijing 100083, China}
\affiliation {Synergetic Innovation Center of Quantum Information and Quantum Physics, University of Science and Technology of China, Hefei, Anhui 230026, China}
\affiliation{College of Materials Science and Opto-Electronic Technology, University of Chinese Academy of Sciences, Beijing 100049, China}
\author{Xiulai Xu}
\affiliation{Beijing National Laboratory for Condensed Matter Physics, Institute of Physics, Chinese Academy of Sciences, Beijing 100190, P. R. China}
\author{Ming Gong}
\thanks{Email: gongm@ustc.edu.cn}
\affiliation{Key Laboratory of Quantum Information, University of Science and Technology of China, CAS, Hefei, 230026, People’s Republic of China}
\affiliation {Synergetic Innovation Center of Quantum Information and Quantum Physics, University of Science and Technology of China, Hefei, Anhui 230026, China}
\author{Shu-Shen Li}
\affiliation{State Key Laboratory of Superlattices and Microstructures, Institute of Semiconductors, Chinese Academy of Sciences,  Beijing 100083, China}
\affiliation {Synergetic Innovation Center of Quantum Information and Quantum Physics, University of Science and Technology of China, Hefei, Anhui 230026, China}
\affiliation{College of Materials Science and Opto-Electronic Technology, University of Chinese Academy of Sciences, Beijing .100049, China}
\author{Guang-Can Guo}
\affiliation{Key Laboratory of Quantum Information, University of Science and Technology of China, CAS, Hefei, 230026, People’s Republic of China}
\affiliation {Synergetic Innovation Center of Quantum Information and Quantum Physics, University of Science and Technology of China, Hefei, Anhui 230026, China}
\date{\today}

\begin{abstract}
The self-assembled quantum dots (QDs) provide an ideal platform for realization of quantum information technology  because it provides {\it on demand}  single photons, entangled photon pairs from biexciton cascade process, single spin qubits, and so on. The fine structure splitting (FSS) of exciton is a fundamental property of QDs for thees applications. From the symmetry point of view, since the two bright exciton states belong to two different representations for QDs with $C_{\text{2v}}$ symmetry, they should not only have different energies, but also have different lifetimes, which is termed exciton lifetime asymmetry. In contrast to extensively studied FSS, the investigation of the exciton lifetime asymmetry is still missed in literature. In this work, we carried out the first investigation of the exciton lifetime asymmetry in self-assembled QDs and presented a theory to deduce lifetime asymmetry {\it indirectly} from measurable qualities of QDs. We further revealed that intrinsic lifetimes and their asymmetry  are fundamental quantities of QDs, which determine the bound of the extrinsic lifetime asymmetries, polarization angles, FSSs, and their evolution under uniaxial external forces.  Our findings provide an important basis to deeply understanding properties of QDs.
\end{abstract}

\pacs{78.67.Hc, 42.50.-p, 73.21.La}
\maketitle

The self-assembled quantum dots (QDs) provide a promising platform for realizing on-demand entangled photon pairs from the biexciton-exciton-vacuum cascade process\cite{Beson00},
which are essential for practical quantum communication\cite{Pan98, Gisin02, Duan01, JSXu10, YFH11, Pan2016}. However the major obstacle in realizing this goal comes from the
non-degeneracy of the two intermediate bright exciton states (see Fig. 1a), in which their energy difference, called fine structure splitting (FSS), is much larger than the homogeneous
broadening of the emission lines ($\Gamma \sim 1$ $\mu$eV\cite{Gammon96,Bayer02,Seguin05}), thus the ''which-way'' information is erased and only classically corrected photons
instead of maximally entangled photon pairs can be created from this cascade process. In the past decade, strenuous efforts have been devoted to eliminate this splitting by applying various experimental techniques, including thermal annealing\cite{Langbein04,Tark04,Ellis07, Seguin, Tartakovskii}, electric field\cite{Gerardot07,Kowalik05,Vogel07, Ghali12, Bennett10, Xiulai1, Xiulai2, Xiulai3}, magnetic field\cite{Stevenson06,Puls99,Stevenson06b, Mrowinski} and external stress\cite{Kuklewicz12, Ding10, Plumhof11, Rastelli12, Trotta12, Trotta, Plumhof13, Kumar2014, Trotta14, Trotta2015, Hofer17} {\it etc.}. However, none of them is efficient. In recent years, the entangled photon pairs were demonstrated in a way that first picking out the QDs with small FSS from a QDs ensemble after post annealing and then eliminating the FSS using magnetic fields (see more details in the first experiment by Stevenson {\it et al.} in 2006\cite{Stevenson06b}). It is worth to note that only a tiny fraction of QDs in experimental grown samples can be used to achieve entangled photon pairs. Moreover,  these devices can only work at low temperature since the emitted exciton energies are very close to the emission lines from the wetting layer\cite{Hafenbrak07}. The mechanism underlying  the difficulty of  eliminating the FSS comes fundamentally from the low symmetry of self-assembled QDs. The higher symmetry of the bulk materials are impossible to be restored in QDs by the above mentioned techniques\cite{Langbein04,Tark04,Ellis07,Gerardot07,Kowalik05,Vogel07,Stevenson06,Puls99,Stevenson06b,Kuklewicz12}. Gong {\it et al.} based on a proposed minimal two-band model  \cite{Gong11} uncovered that the FSS is impossible to be tuned to zero for a general QD with a single external force. However, the FSS can be eliminated by two independent external forces\cite{Gong13}. Employing three external forces, it is even possible to construct a wavelength tunnable entangled photon emitter\cite{WangJP12, WangJP15, Trotta15}, opening the door for interfacing between QDs and other solid state systems and even between dissimilar QDs.  Recently, this proposal has indeed  been realized in experiments \cite{ZhangJX2015,ChenYan2016,ZhangJX2016,ZhangJX2013, HuangH17}, in which the wavelength of exciton can be tuned in the range of few meV.

\begin{figure}
\centering
\includegraphics[width=3.3in]{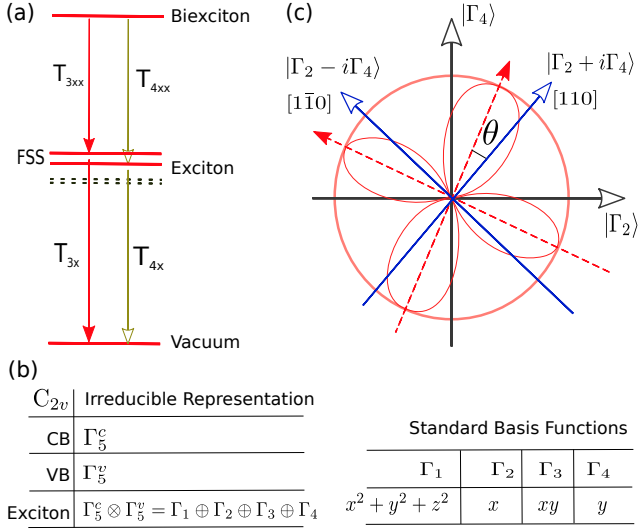}
    \caption{(Color online) Direct consequence of low symmetry in QDs. (a) The two bright states (solid lines) of exciton have different energies and lifetimes when they belong to different irreducible
    representations. (b) The basic irreducible representations of conduction band (CB) and valance band (VB) and exciton, and their basis for QDs with $C_{\text{2v}}$ symmetry. (c)
    In the presence of weak $C_1$ potential, the direct coupling between the two bright states lead to wave function mixing, thus the emission will deviate from $[110]$ and $[1\bar{1}0]$ directions.}
\label{fig-fig1}
\end{figure}

From the symmetry point of view, since the two bright exciton states belong to two different representations for QDs with $C_{\text{2v}}$ symmetry\cite{Koster63}, they should not only have different energies, but also have different lifetimes. The lifetime difference between two bright exciton states are termed as exciton lifetime asymmetry (see Fig. 1a). In contrast to extensively studied FSS, the investigation of this anisotropy effect is still missed in literatures. In this work, we carried out the first investigation of the exciton lifetime asymmetry in self-assembled QDs and presented a theory to deduce lifetime asymmetry {\it indirectly} from measurable qualities of QDs. We further revealed that the lifetime asymmetries are fundamental quantities of QDs and unraveled some exact relations between FSS, polarization angle and lifetime asymmetry for exciton and biexciton in QDs. These exact relations are verified by performing atomistic simulations of self-assembled QDs using the empirical pseudopotential method\cite{Wang99, Wang00}.

{\it Atomistic Simulation Method}. We employ the empirical pseudopotential method\cite{Wang99, Wang00} to simulate the electronic and optical properties of self-assembled QDs. We model the InGaAs/GaAs QDs by embedding it a much larger GaAs supercell with periodic boundary condition and minimize the total strain energy using the valence force field
model\cite{Keating66,Martin70}. The single particle wave functions are determined by,
\begin{equation}
(-{1\over 2}\nabla^2 + V_{\text{ps}}({\bf r}))\psi_{i} = E_{i} \psi_{i},
\end{equation}
where $V_{\text{ps}} = \mathcal{V}_{\text{soc}} + \sum_{i\alpha} \mathcal{V}_{i\alpha}({\bf r - R}_{i\alpha})$ is the empirical pseudopotential and $i$ is the site index,  ${\bf R}_{i\alpha}$ is the position of the atom type $\alpha$, and $\mathcal{V}_{\text{soc}}$ the spin-orbit coupling term. The position of each atom in the supercell is obtained by minimizing the total strain energy. The exciton and biexciton energies are then calculated by employing the configuration
interaction method taking into account the Coulomb interaction and exchange-correlation interaction \cite{Wang99, Wang00, Franceschetti99}.

{\it Theoretical Modelling}. We carry out the symmetry analysis of QDs as given in Ref. \onlinecite{Gong11} by decomposing the QD Hamiltonian $H$ into two parts: $H = H_{\text{2v}} + V_1$, where $H_{\text{2v}}$ is a predominant term having a $C_{\text{2v}}$ symmetry and includes the kinetic energy, Coulomb interactions and all the potentials with $C_{\text{2v}}$ symmetry.  The second term $V_1$ represents the remaining potentials lowering the symmetry of QDs from $C_{\text{2v}}$ to $C_1$. $V_1$ can be treated as a perturbation to $H_{\text{2v}}$ considering its weak effect on energy levels of QDs. Two bright states of the $H_{\text{2v}}$ Hamiltonian are denoted as $|3\rangle = |\Gamma_2 - i\Gamma_4\rangle$ and
$|4\rangle = |\Gamma_2 + i\Gamma_4\rangle$, respectively,  where $|\Gamma_i\rangle$ ($i=1,2,3,4$) are the irreducible representations of the $C_{\text{2v}}$ point group (see Fig. \ref{fig-fig1}b and the symmetry table in \onlinecite{Koster63}). We refer quantities of the $H_{\text{2v}}$ Hamiltonian to intrinsic. Since these two bright exciton states belong to two different irreducible representations, they must have different energies with an energy separation of FSS, and different lifetimes ($\tau_{3\eta}$ and $\tau_{4\eta}$, respectively) with a time difference termed as intrinsic lifetime asymmetry $\delta \tau_\eta$. Hence,  $\tau_{3\eta} = \tau_{\eta} + \delta \tau_\eta/2$ and $\tau_{4\eta} = \tau_\eta - \delta \tau_\eta/2$; here, $\tau_\eta$ is averaged lifetime (hereafter $\eta=\text{x}$ for exciton and $\eta=\text{xx}$ for biexciton). Since the $V_1$ potential is inevitable and uncontrollable in experimental grown QDs, the intrinsic lifetimes and their asymmetry  can not be measured  {\it directly}  in experiments.

To simplify the effective model for total QD Hamiltonian $H$, we take advantage of two additional features. Firstly, the time-reversal symmetry for exciton, $\mathcal{T}^2 = +1$ (excitons have integer spin), ensures that the wave functions of two bright exciton states can be real simultaneously\cite{Sakurai}. Secondly, the spin selection rule forbids the mixture between dark exciton states ($m = \pm 2$) and bright exciton states ($m = \pm 1$) even in the presence of $V_1$ term.
The dark exciton states can only be probed by coupling to bright states through applying external (in-plane) magnetic fields\cite{Stevenson06,Puls99,Stevenson06b}. We thus can safely neglect the dark states and construct the effective model based only on  two bright exciton states of the $H_{\text{2v}}$ as following,
\begin{equation}
H = E_0 + \delta \sigma_z + \kappa \sigma_x
    \label{eq-H}
\end{equation}
where $E_0$ is the mean energy of two bright exciton states, $\sigma_{x}$ and $\sigma_z$ are Pauli matrices acting on two bright states with eigenvalues of $m = \pm 1$, and $\delta = \langle 3|H|3\rangle - \langle 4|H|4\rangle$
and $\kappa = \langle 3|H|4\rangle$. The bright states of the total Hamiltonian $H$ can be constructed as a linear combination of $|3\rangle$ and $|4\rangle$,
\begin{equation}
\begin{pmatrix}
\psi_3 \\
\psi_4
\end{pmatrix}
= u(\theta)
\begin{pmatrix}
|3\rangle \\
|4\rangle
\end{pmatrix}, \quad
u(\theta) =
\begin{pmatrix}
\cos(\theta)  & \sin(\theta)  \\
-\sin(\theta)  & \cos(\theta)
\end{pmatrix},
\label{eq-psi}
\end{equation}
where $\theta$ is the polarization angle as schematic shown in Fig. 1c and $u(\theta)$ is a so(2) rotation matrix. We learn that $\tan(\theta) = (\delta + \Delta/2)\kappa^{-1}$, where $\Delta = 2\sqrt{\delta^2 + \kappa^2}$  the magnitude of exciton FSS. The values of $\kappa$ and $\delta$ can be determined from experimental measurements of FSS and polarization angle. In an ensemble of QDs, the $V_1$ potential is usually a random potential, thus $\kappa$ and $\delta$ can be treated as two independent random numbers\cite{Gong14}, for which reason the QDs with similar structural profile may have fairly different optical properties.

\begin{table*}[t]
    \centering
    \caption{{\bf Summarized parameters for different QDs from atomistic simulation}. We consider the lens (L), elongated (E) and pyramind (Py) QDs with different sizes (diameter
    sizes along $[110]$ and $[1\bar{1}0]$ directions and height along $z$ direction). The exciton energy, FSS and polarization angle without external force are shown in columns fourth to sixth. The
    last eight columns present the extrinsic and intrinsic lifetime asymmetries. }
    \begin{tabular}{p{0.02\textwidth} p{0.13\textwidth} p{0.12\textwidth} p{0.05\textwidth} p{0.055\textwidth} p{0.05\textwidth} p{0.059\textwidth} p{0.059\textwidth} p{0.059\textwidth} p{0.059\textwidth} p{0.059\textwidth} p{0.059\textwidth} p{0.059\textwidth} p{0.059\textwidth} } \hline\hline
        \#  & QD(P/L)   & ($d_{[110]}$, $d_{[1\bar{1}0]}$, $h$) & $E_\text{X}$(eV)   & $\Delta$($\mu$eV)  & $\theta$  & $T_\text{x}$(ns)   & $\delta T_\text{x}$(ps)  &  $\tau_{\text{x}}$(ns)  & $\delta\tau_{\text{x}}$(ps)  & $T_\text{xx}$(ns)   & $\delta T_\text{xx}$(ps)  &  $\tau_{\text{xx}}$(ns)  & $\delta\tau_{\text{xx}}$(ps) \\ \hline
1   & InAs(L)     & 25.0, 25.0, 3.0       &  1.01  &  14.7   &  0  &  1.56  &  134.00  &  1.56  &  134.00  & 1.67 &  144.45  &  1.67  &  144.45   \\
2   & InAs(L)     & 24.0, 24.0, 3.0       &  1.02  &  16.3   &  0  &  1.56  &  142.77  &  1.56  &  142.77  & 1.67 &  153.65  &  1.67  &  153.65   \\
3   & InAs(L)     & 25.0, 25.0, 4.0       &  0.98  &   8.2   &  0  &  1.84  &  172.55  &  1.84  &  172.55  & 1.96 &  184.84  &  1.96  &  184.84   \\
4   & InAs(L)     & 25.0, 20.0, 2.5       &  1.02  &  6.0    &  0  &  1.56  &  385.34  &  1.56  &  385.34  & 1.68 &  418.26  &  1.68  &  418.26   \\
5   & InAs(L)     & 25.0, 22.7, 3.0       &  0.99  &  6.1    &  0  &  1.59  &  244.85  &  1.59  &  244.85  & 1.71 &  265.04  &  1.71  &  265.04   \\
6   & In$_{0.6}$Ga$_{0.4}$As(L)  & 24.0, 24.0, 4.0 & 1.24  & 1.29  &  14.65  & 1.21  &  -1.48  &  1.21  & -1.70  &  1.33  &  -2.52  &  1.33  &  -2.89  \\
7   & In$_{0.6}$Ga$_{0.4}$As(L)  & 25.0, 25.0, 3.0 & 1.25  & 1.35  & 174.52 & 1.29  &  15.52 &  1.29  & 15.81 &  1.40  &  18.1  &  1.40  &  18.45  \\
8  & In$_{0.6}$Ga$_{0.4}$As(L)  & 25.0, 25.0, 4.0 & 1.23  & 3.37  & 157.68 & 1.23  &  6.87  &  1.23  & 9.65  &  1.33  &  8.15  &  1.33  &  11.45  \\
9  & In$_{0.6}$Ga$_{0.4}$As(E)  & 25.0, 22.7, 3.0 & 1.24  & 7.48  & 122.35 & 1.23  & -46.45  &  1.23  & 108.85  &  1.35  &  -52.97   & 1.35  &   124.14  \\
10  & In$_{0.6}$Ga$_{0.4}$As(E)  & 25.0, 30.0, 4.5 & 1.21  & 4.23  & 109.16 & 1.13  & -147.54 &  1.13  & 188.57  &  1.26  &  -166.49  & 1.26  &  212.80  \\
11  & In$_{0.6}$Ga$_{0.4}$As(Py)  & 25.0, 25.0, 3.0 & 1.28  & 7.18  & 163.30 & 1.43  & -3.0 &  1.43  & -3.59  &  1.53  &  -2.91  &  1.53  &  -3.49  \\
12  & In$_{0.6}$Ga$_{0.4}$As(Py)  & 25.0, 25.0, 4.0 & 1.25  & 6.44  & 164.01 & 1.30  & -8.64 &  1.30  & -10.19  &  1.41  &  -11.62  &  1.41  &  -13.70  \\
\hline
    \end{tabular}
    \label{tableI}
\end{table*}

If two bright exciton states have measurable extrinsic lifetimes $T_\text{3$\eta$}$ and $T_\text{4$\eta$}$, which are usually deduced experimentally by fitting the time-resolved photoluminescence spectrum to a single exponential decaying function. The difference in lifetimes gives rise to the extrinsic lifetime asymmetry $\delta T_{\eta}$,
\begin{equation}
    \delta T_{\eta} = T_{3\eta} - T_{4\eta}, \quad T_{3\eta} + T_{4\eta} = 2T_{\eta},
    \label{eq-defTx}
\end{equation}
where $T_{\eta}$ is the mean extrinsic lifetime measured from off-resonant excitation. In this measurement, we have assumed that the spin information in the electron-hole pairs are totally lost during the relaxation from the wetting layer to the ground state of exciton, thus the two bright states are equally populated. However, those extrinsic lifetime asymmetries are not well defined for following reasons. Firstly, the time-resolved photoluminescence spectrum is in fact composed by two exponential decay functions with two slightly different lifetimes\cite{Mukai1996, Inokuma90, Wilfried2002}, and thus $T_{3\eta}$ and $T_{4\eta}$ can only be obtained in the sense of best fitting. Secondly, the $V_1$ potential, which induces inter-state mixing (see Eq. \ref{eq-psi}), is not the major origin of lifetime asymmetries \cite{Singh09, Dup11}. A more accurate description of lifetime asymmetries should be defined by the model of $H_{\text{2v}}$, instead of the total Hamiltonian $H$.

Because lifetime asymmetries are in general much smaller than the mean lifetimes, we are ready to obtain
\begin{equation}
    {1\over T_{3\eta}}= {\cos^2(\theta) \over \tau_{3\eta}} + {\sin^2(\theta) \over \tau_{4\eta}},
    {1\over T_{4\eta}}= {\sin^2(\theta) \over \tau_{3\eta}} + {\cos^2(\theta) \over \tau_{4\eta}}.
    \label{eq-per}
\end{equation}
To the leading term of $\delta \tau_\eta$,
\begin{equation}
    T_{3\eta} = \tau_\eta + {\cos(2\theta) \over 2} \delta \tau_\eta, \quad T_{4\eta} = \tau_\eta - {\cos(2\theta) \over 2} \delta \tau_\eta.
    \label{eq-T34}
\end{equation}
The above results are identical to fitting the photoluminescence spectrum to a single exponential decaying function by minimizing the following functional,
\begin{equation}
    \mathcal{F}_n = \int_0^{\infty} | c_n e^{-{t \over \tau_{3\eta}}} + (1-c_n) e^{-{t \over \tau_{4\eta}}} - e^{-{t \over T_{\eta}}}|^2 dt,
    \label{eq-F}
\end{equation}
where $n=3, 4$, with $c_3 = \cos^2(\theta)$ for $|3\rangle$ and $c_4 = \sin^2(\theta)$ for $|4\rangle$.  Assuming $T_{\eta} = \tau_\eta + x \delta \tau_\eta$ (where $|x| \ll 1$), we obtain $\mathcal{F}_{\eta} = (-1 + 2c_n + 2x)^2 \delta \tau^2/16\tau + \mathcal{O}(\delta \tau/\tau)^3$. It is straightforward to obtain its solution, which is identical to ones given in Eq. \ref{eq-T34}.  We therefore see that definitions given in Eq. \ref{eq-per} and Eq. \ref{eq-F} are equivalent  in the small asymmetry limit.

\paragraph{The unusual effects of low symmetry perturbation potential.}
(i) Lifetime sum rule: The weak $C_1$ potential will not alter the averaged exciton lifetime, which is determined as,
\begin{equation}
    T_{3\eta} + T_{4\eta} = \tau_{3\eta} + \tau_{4\eta}, \quad \tau_\eta =T_\eta.
    \label{eq-sum}
\end{equation}
The equality relation $\tau_\eta = T_\eta$ indicates that the mean lifetime is independent of $V_1$ potential, which is manifested in the investigated QDs ensembles as shown in Fig. \ref{fig-fig3}. (ii) Lifetime asymmetry relation: The extrinsic lifetime asymmetry is determined as
\begin{equation}
    \delta T_\eta = T_{3\eta} - T_{4\eta} = \cos(2\theta)\delta \tau_\eta \le |\delta \tau_{\eta}|.
    \label{eq-cos2theta}
\end{equation}
We see that, to the leading term, the extrinsic lifetime asymmetries depends only on the intrinsic lifetime asymmetries $\delta \tau_{\eta}$ and the polarization angle $\theta$, and is independent of the mean lifetimes $\tau_{\eta}$ and $T_{\eta}$. Interestingly, while the $V_1$ potential can enhance the FSS and polarization angle, it will unexpectedly suppress the magnitude of $\delta T_{\eta}$, which is upper bounded by $|\delta \tau_{\eta}|$. Moreover, when $\delta \tau_\eta = 0$ (in QDs with high symmetries) or $\theta = \pm {\pi \over 4}$ (polarized along $\Gamma_2$ and $\Gamma_4$ directions), the change of low symmetry potential will never induce a finite extrinsic lifetime asymmetry.

\begin{figure}
    \centering
    \includegraphics[width=3.2in]{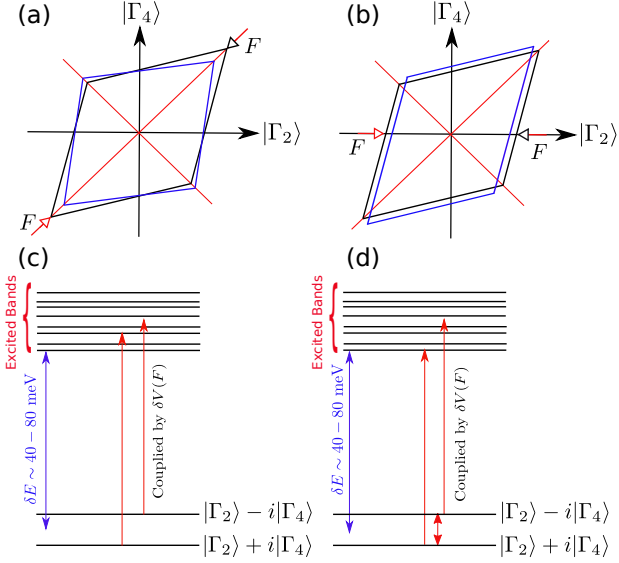}
    \caption{(Color online) Effect of external stress on QDs. (a) and (b) show the role of stress along $[110]$ and $[100]$ direction, respectively. The black and blue
    boxes represent the QDs before and after being stressed. (c) and (d) show the coupling between the bright states and the excited states coupled due to stress along different directions. In these two cases the strain Hamiltonian may have different symmetries, thus the coupling between the bright states and excited states and the direction coupling among the two bright states are totally different. }
    \label{fig-fig2}
\end{figure}

\begin{figure}
    \centering
    \includegraphics[width=3.4in]{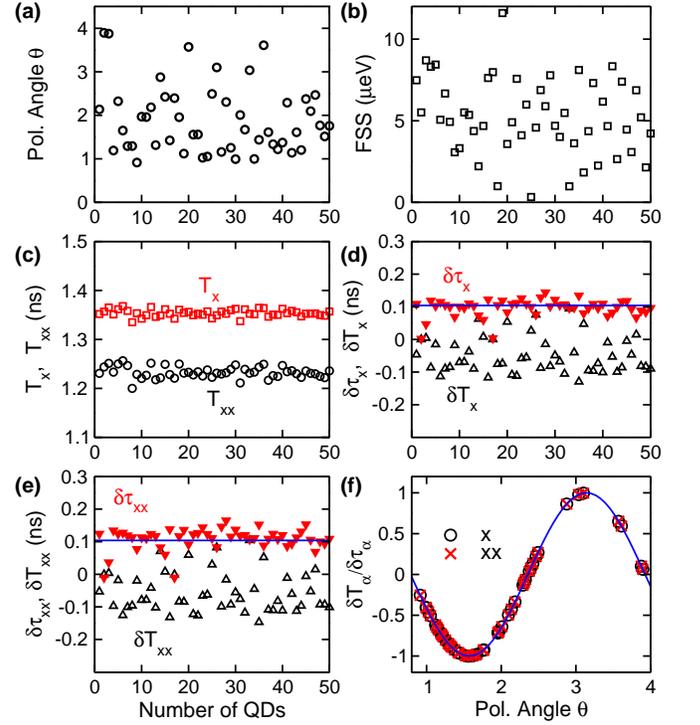}
    \caption{(Color online) Effect of random fluctuation the lifetime asymmetries in QDs ensemble. (a) and (b) show the calculated polarization angles and FSSs in QDs ensemble. (c) The extrinsic
    mean lifetimes for exciton and biexciton (by assuming Eq. \ref{eq-sum}). (d) - (e) The intrinsic and extrinsic lifetime asymmetries in QDs ensemble for
    exciton and biexcition, where the intrinsic lifetime asymmetries are determined using the wave functions of $|3\rangle$ and $|4\rangle$ for QDs with $H_{\text{2v}}$ symmetry. The horizontal
    blue lines represent the mean value of $\delta \tau_\eta$, which are 0.10 ns ($\sim 9\% \tau_\text{x}$) and 0.11 ns ($\sim 10\% \tau_\text{xx}$), respectively.
    (f) The ratio between $\delta T_\eta$ and $\delta \tau_\eta$ will collapse to a cosine function (blue solid line)  according to Eq. \ref{eq-cos2theta}.}
    \label{fig-fig3}
\end{figure}

To verify above predictions we carry out atomistic simulations for single QDs as well as QDs ensembles. The calculated results for various types of single pure InAs/GaAs QDs and alloyed In(Ga)As/GaAs QDs are summarized in Table \ref{tableI}. For pure  InAs/GaAs QDs, we find that, as expected since  $V_1=0$, the polarization angle $\theta = 0$ (or $\pi/2$), and $\delta \tau_\eta = \delta T_\eta$,  following exactly  Eq. \ref{eq-cos2theta}. The magnitude of both intrinsic and extrinsic lifetime asymmetries $\delta T_{\eta}$ and $\delta \tau_{\eta}$ are in range of 0.1 - 0.4 ns, and much smaller than the mean lifetimes $T_{\eta}$ and $\tau_{\eta}$, which are around 1.5-2.0 ns.
From Table I we see that the lifetime sum rule given in Eq . \ref{eq-sum} holds for all calculated QDs, including alloyed ones. Although the polarization angles $\theta$ in alloyed QDs deviate significantly away from the [110] and $[1\bar{1}0]$ directions caused by wave functions mixing,  $|\delta T_\eta| \le |\delta \tau_\eta|$ holds in all QDs. Therefore, we demonstrate that the low symmetry perturbation potential
can remarkably suppress the lifetime asymmetries along with the derived relation  given in Eq. \ref{eq-cos2theta}.

\begin{figure}
    \centering
    \includegraphics[width=0.48\textwidth]{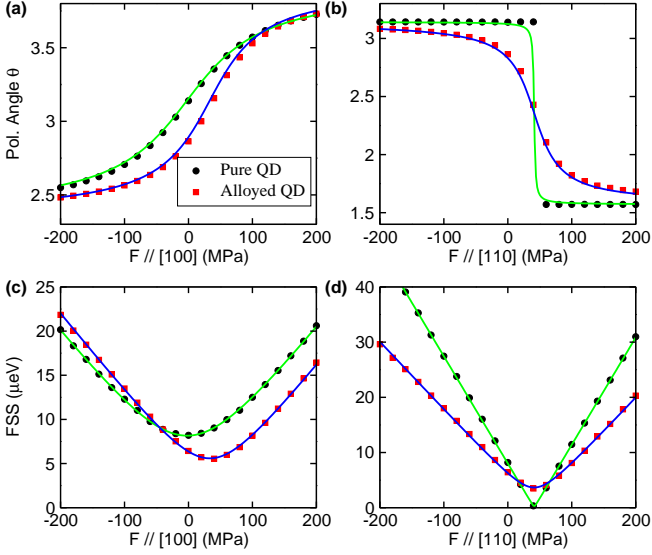}
    \caption{(Color online) Polarization angle and FSS in pure and alloyed QDs under stress. The left (a, c) and right (b, d) columns show the results for stress along [100] direction and $[110]$ direction, respectively. The open symbols represent the data from atomistic simulation while the solid line is the fitting using the effective two-band model with the following parameters in unit of $\mu$eV for $\delta, \kappa$ and
    $\mu$eV/MPa for $\alpha, \beta$: (i) For pure QDs, $\delta$=2.73, $\kappa$=-1.69, $\alpha_{[100]}$=0, $\beta_{[100]}$=0.05 and
    $\alpha_{[110]}$=-0.12, $\beta_{[110]}$=0. (ii) For alloyed QDs, $\delta$=4.1, $\kappa$=-0.02, $\alpha_{[100]}$=0, $\beta_{[100]}$=0.05 and $\alpha_{[110]}$=-0.19, $\beta_{[110]}$=0.}
    \label{fig-fig4}
\end{figure}

We further consider the alloyed QDs ensembles, in which the FSSs, polarization angles, and mean lifetimes $T_\eta$ fluctuate from dot to dot in a wide range. We arbitrarily choose an alloyed InGaAs/GaAs QD from Table \ref{tableI} (No. 9) and then generate 50 different replica with randomly placed In and Ga atoms for a specific composition of 60\%, which mimics an experimentally grown QDs ensemble. The calculated results for this ensemble are shown in Fig. \ref{fig-fig3}. As expected, the FSSs, polarization angles $\theta$, mean lifetimes $T_\eta$ as well as lifetime asymmetries $\delta T_\eta$  fluctuate from dot to dot in a wide range within an ensemble. Because all modeled dots within the ensemble have the same shape, size and alloy composition, they share always the same $H_{\text{2v}}$ Hamiltonian, which means their intrinsic properties should be similar. Indeed, the deduced intrinsic lifetime asymmetries $\delta \tau_\eta$  are rather insensitive to alloy atom fluctuations in an ensemble. The observed fluctuations in extrinsic quantities of FSSs, $\theta$, $T_\eta$, and $\delta T_\eta$ are attributed to alloy fluctuation induced change of the $V_1$ potential.  We find that $|\delta T_\eta| \le |\delta \tau_\eta|$ for all dots (see Fig. \ref{fig-fig3}d-e), and the ratios between extrinsic and intrinsic lifetime asymmetries fall on a cosine curve as shown in Fig. \ref{fig-fig3}f, in  consistent with the prediction given in  Eq. \ref{eq-cos2theta}. The alloy-induced $V_1$ perturbation potential remarkably reduces extrinsic lifetime asymmetries $\delta T_\eta$ to around zero in compared with their intrinsic lifetime asymmetry around 0.1 ns.

\begin{figure}
    \centering
    \includegraphics[width=0.47\textwidth]{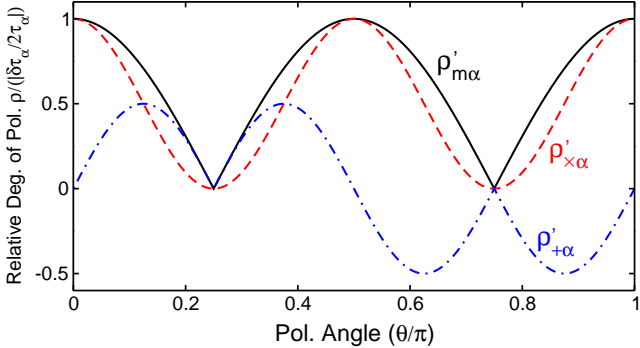}
    \caption{(Color online) Relative relations between different degree of polarizations as a function polarization angle for an alloyed quantum dots. }
    \label{fig-fig5}
\end{figure}

\paragraph{The optical anisotropy relying on the intrinsic lifetimes.} The signals collected along $\phi$ direction after off-resonance excitation can be written as
\begin{equation}
    I_{\eta}(\phi) \propto 1 + \cos(2\theta) \cos(2(\phi-\theta)) {\delta \tau_\eta \over 2\tau_\eta},
    \label{eq-signal}
\end{equation}
where $\phi$ is the angle relative to $[110]$ direction. Here, we also assume that both bright states are equally occupied from off resonance excitation like we developing Eq. \ref{eq-sum}.
 We gain the maximum degree of linear polarization $\rho_{\eta, {\text{max}}}$  when $\phi = \theta$ or $\theta + \pi/2$,
\begin{equation}
    \rho_{\eta, {\text{max}}} = {I_{\text{max}} - I_{\text{min}}  \over I_{\text{max}} + I_{\text{min}}} = \Big|{\cos(2\theta) \delta \tau_\eta \over 2\tau_\eta}\Big| \le \Big|{\delta \tau_\eta \over 2\tau_\eta} \Big |.
\label{eq-lp}
\end{equation}
We learn that the degree of polarization is determined fully by polarizaton angle $\theta$, lifetime  asymmetry $\delta \tau_\eta$, and mean lifetime $\tau_\eta$, considering that the transition between two bright exciton states are strictly forbidden by selection rule. The degree of polarization is upper bounded by $|\delta \tau_\eta/\tau_\eta|/2$. Regarding $|\delta \tau_\eta| \ll \tau_\eta$, the degree of polarization is restrict to small magnitudes despite the large varying from dot to dot. Note that the polarization discussed here should different from the linear polarization defined in some of experiments, in which the excitation and measurement are performed along two orthogonal direction, thus the degree of polarizations are generally in
the order of 80\% - 90\% out of the available experimental data\cite{Kulakovskii1999,Ulrich2003,Astakhov2006,Dzhioev1998} . In the latter case polarization is complicated since carrier scattering, spin flipping and lifetime asymmetry all contribute to its non-unity.

For the measurement along $[110]$ and $[1\bar{1}0]$ directions  (let $\phi = 0$),
\begin{equation}
    \rho_{\eta, \times} = {I_{[110]} - I_{[1\bar{1}0]}  \over I_{[110]} + I_{[1\bar{1}0]}} = {\cos^2(2\theta) \delta \tau_\eta  \over 2\tau_\eta},
\end{equation}
which is smaller than $\rho_{\eta,\text{max}}$ by a factor of $\cos(2\theta)$. We further prove straightforwardly that the degree of linear polarization along
the $x$ and $y$ directions is $\rho_{\eta,\text{+}} = \sin(2\theta)\rho_{\eta,\text{max}}$. The relations between these degree of polarizations are presented in Fig. \ref{fig-fig5}.
Although the small extrinsic lifetime asymmetry maybe hard to measure directly by fitting to two exponential decay functions, the relative polarization angle $\theta$, the mean lifetime $\tau_\eta$ (see Eq. \ref{eq-sum}) as well as the degree of polarization $\rho_{\eta, i}$ can be measured precisely, thus making the intrinsic lifetime asymmetries $\delta \tau_\eta$ rather feasible to get in experiments. Once we obtain intrinsic lifetime asymmetry $\delta \tau_\eta$, the extrinsic lifetime asymmetry $\delta T_\eta$   can be instead deduced using Eq. \ref{eq-cos2theta}.

\begin{figure}
    \centering
    \includegraphics[width=0.48\textwidth]{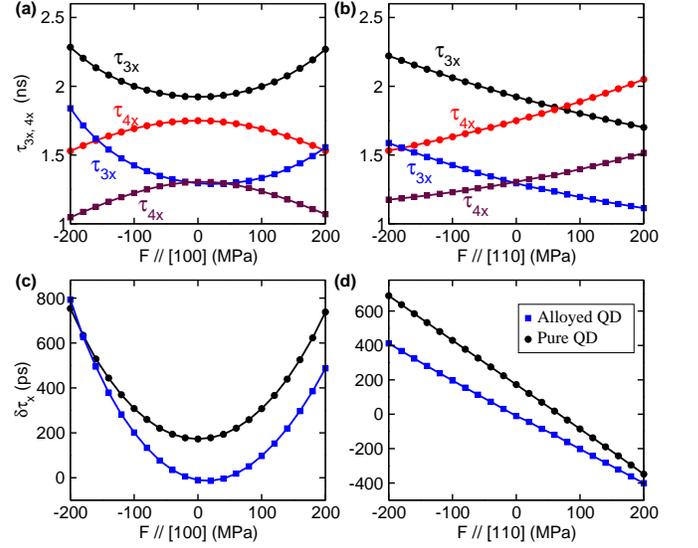}
    \caption{(Color online) Linear and quadratic relations for intrinsic lifetime asymmetries. The solid squares show the pure QDs, while the solid circles show the alloyed QDs along [100] (left
        column) and [110] (right column) direction obtained from atomistic simulation. The solid lines are fitted using quadratic function of $F$, and in the solid lines we have used the parameters
        (in unit of ns/MPa for $\xi_i^\text{x}$ and ns/MPa$^2$ for $\gamma_i^\text{x}$) as follows (i) For pure QDs along [100] direction: $\gamma_3^{\text{x}}$=6.73$\times$$10^{-6}$, $\xi_3^{\text{x}}$=-2.29$\times$$10^{-6}$,  $\gamma_4^{\text{x}}$=7.53$\times$$10^{-7}$, $\xi_4^{\text{x}}$=1.83$\times$$10^{-6}$; For pure QDs along [110] direction: $\gamma_3^{\text{x}}$=3.52$\times$$10^{-4}$, $\xi_3^{\text{x}}$=-2.31$\times$$10^{-8}$, $\gamma_4^{\text{x}}$=-4.20$\times$$10^{-4}$, $\xi_4^{\text{x}}$=-2.30$\times$$10^{-8}$. (ii) For alloyed QDs along [100] direction: $\gamma_3^{\text{x}}$=3.45$\times$$10^{-4}$, $\xi_3^{\text{x}}$=-5.53$\times$$10^{-6}$, $\gamma_4^{\text{x}}$=-3.51$\times$$10^{-5}$, $\xi_4^{\text{x}}$=3.82$\times$$10^{-6}$; For alloyed QDs along [110] direction: $\gamma_3^{\text{x}}$=7.00$\times$$10^{-4}$, $\xi_3^{\text{x}}$=-2.18$\times$$10^{-7}$, $\gamma_4^{\text{x}}$=-4.95$\times$$10^{-4}$, $\xi_4^{\text{x}}$=-2.58$\times$$10^{-7}$.  }
    \label{fig-fig6}
\end{figure}

\paragraph{The optical polarization relying on the intrinsic  lifetimes} Since exciton and biexciton states possess the same polarization angle $\theta$, we obtain
\begin{equation}
    {\rho_{\text{x},i} \over \rho_{\text{xx},i}} = {\delta \tau_\text{x} \over \delta \tau_{\text{xx}}} \cdot {\tau_{\text{xx}} \over \tau_\text{x}}, \quad
    i = \{\text{max}, \times, +\},
\end{equation}
which is also independent of the measured $\phi$. The deduced relations of the degree of polarizations with intrinsic  lifetimes $\tau_\eta$ and lifetime asymmetry $\delta \tau_\eta$ of the $H_{\text{2v}}$ term along various directions can be examined with the data of atomistic calculation shown in Fig. \ref{fig-fig3}. For this QDs ensemble, $\delta \tau_\eta \sim 0.1$ ns, $\tau_\eta \sim 1.25 - 1.35$ ns, thus $\rho_{\eta, \text{max}}\leqslant \delta \tau_\eta/2\tau_\eta\sim 4\%$, which is in good agreement with atomistic simulation predicted upper bound of the degree of polarization of $\sim 4\%$ as shown in Fig. 3a. We may also calculate the ratio between the degree of polarizations for exciton and biexciton, which is independent of directions of optical measurements. We find this ratio being $\rho_{\text{x},i}/\rho_{\text{xx},i} \sim 1.08$. We expect this ratio becomes larger for QDs with  large shape asymmetry (such as  elongation). In principle, the evolution of linear polarization as a function of angle $\phi$ (Eq. \ref{eq-signal}) can be obtained by carefully calibrating the light path in experiments.

\paragraph{The Quadratic and Strong Nonlinear responses of the extrinsic lifetimes to external stress.}
In the presence of weak external force $F$, the effective Hamiltonian becomes\cite{Gong11, Gong13, WangJP12, WangJP15},
\begin{equation}
    H = E_0 + (\alpha F + \delta)\sigma_z + (\kappa + \beta F) \sigma_x,
    \label{eq-HF}
\end{equation}
where, the extra perturbative term $\delta V = V_s F$ is responsible for the applied external force. The symmetry of $\delta V$ depends strongly on the directions and the ways the forces been applied (see Fig. \ref{fig-fig2}). Here, we only consider the case with single external force. The additivity of $\delta V$ ensures that multiple forces can be treated in the same way as the single force.
Notice that two unperturbed (intrinsic) bright exciton states $|3\rangle$ and $|4\rangle$  are also functions of $F$ due to the $H_{\text{2v}}$ Hamiltonian depends on $F$.  For the stress applied along the $[110]$ direction, the strained Hamiltonian still keeps the $C_{\text{2v}}$ symmetry (see Fig. \ref{fig-fig2}a, c), thus the stress induced coupling between the two bright states is forbidden. This is different from the case along the $[100]$ direction, where the stress induces not only coupling of bright exciton states to highly excited states arising from other bands (termed as inter-band coupling),  but also coupling between two bright exciton states (termed as intra-band coupling) (see Fig. \ref{fig-fig2}b, d). The linear coupling dominates the inter-band coupling due to the large energy separation between the ground $s$ band and the excited $p$ and $d$ bands, while the nonlinear effect may become significant in intra-band coupling due to much smaller energy difference between these two bright exciton states. This difference has remarkable consequences to the lifetime asymmetries. According to perturbation theory, we rewrite the intrinsic lifetimes as,

\begin{equation}
    {1 \over \tau_{i}^{\eta}(F)} = {1 \over \tau_i^{\eta}} + \gamma_i^\eta F + \xi_i^\eta F^2,
\end{equation}
for $i = 3, 4$ and $\eta = \{\text{x}, \text{xx}\}$. Here, we have introduced $\gamma_i^\eta$ and $\xi_i^\eta$ to characterize the linear and quadratic dependence on external force, respectively. When the contributions of the second and third terms are small in compared with $1/\tau_i^{\eta}$, we obtain,
\begin{equation}
    \tau_i^\eta(F) = \tau_i^{\eta} - (\tau_i^{\eta})^2 [ \gamma_i^\eta F + \xi_i^\eta F^2 - (\gamma_i^\eta)^2\tau_i^{\eta} F^2],
\end{equation}
which is again a quadratic function of $F$. It it expected that $\tau_\eta = (\tau_3^\eta + \tau_4^\eta)/2$ and $\delta \tau_\eta = (\tau_3^\eta - \tau_4^\eta)/2$ are also linear and/or quadratic functions of $F$. With these defined intrinsic lifetime asymmetries, the extrinsic lifetimes and their asymmetries can be obtained via,
\begin{equation}
    \delta T_\eta(F) = \cos(2\theta(F)) \delta \tau_\eta(F), \quad T_\eta(F) = \tau_\eta(F).
    \label{eq-deltaTF}
\end{equation}
The striking consequence is that even under a weak force the extrinsic lifetime asymmetry $\delta T_\eta$ is not necessary to be a simple quadratic function of $F$ due to the presence of the strong nonlinear cosine term. We can derive  more exact relations via a combination of current results and results in previous literatures\cite{Gong11, WangJP12, WangJP15}.

We next attempt to study the evolution of optical properties of QDs under external forces, which enables us to verify the last and the most intriguing prediction (Eq. \ref{eq-deltaTF}) in this work. The properties of QDs fluctuate strongly from dot to dot, therefore the investigation in single QDs could provide more convincing evidences for our predictions. We consider the pure InAs/GaAs QD (No. 3 in Table I) and alloyed InGaAs/GaAs QD (No. 12) under uniaxial stress along [110] and [100] directions, respectively. Fig. \ref{fig-fig4} shows calculated FSSs and polarization angles as functions of stress for these two QDs employing the atomistic method, accompanying fitted results using the two-level model. We demonstrate that we can tune the FSS to zero upon applying a stress along the [110] direction, which does not lower the QD symmetry (see Fig. \ref{fig-fig2}), for the pure QDs, but it is impossible for alloyed QDs where the achievable minimum lower bound of FSS is significant larger than the spontaneous broadening of the spectra  $\Gamma$\cite{Gammon96,Bayer02,Seguin05}. This large minimum FSS implies that we can not eliminate the FSS of alloyed QDs by using one single external force.

\begin{figure}
    \centering
    \includegraphics[width=0.48\textwidth]{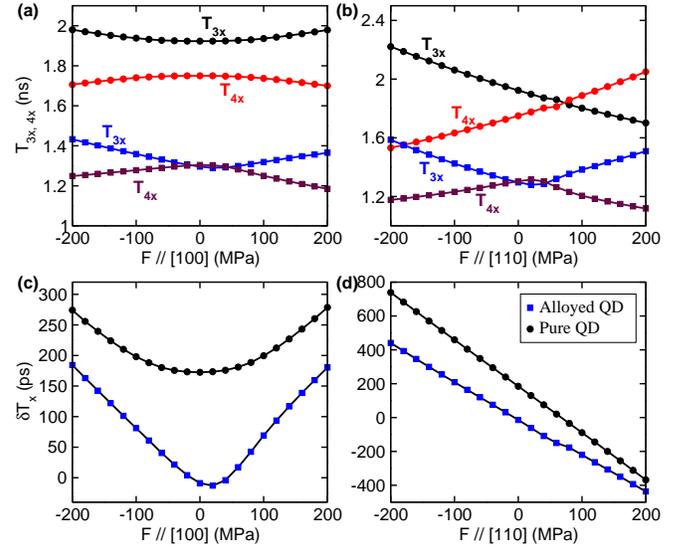}
    \caption{(Color online) Strong nonlinearity in extrinsic lifetimes and extrinsic lifetime asymmetries. The solid squares, and solid angles show the results for alloyed and pure QD, respectively.
    The solid line is computed using parameters from Fig. \ref{fig-fig6}, and using Eq. \ref{eq-deltaTF}.}
    \label{fig-fig7}
\end{figure}

Fig. \ref{fig-fig6} shows calculated intrinsic lifetimes and their asymmetries as functions of applied stress for both pure InAs/GaAs and alloyed InGaAs/GaAs QDs. For stress applied along the $[110]$ direction, we find that $\tau_{\text{3x}}$ and $\tau_{\text{4x}}$ are linear functions of stress $F$, as shown in Fig. \ref{fig-fig6}b-d. To further verify their linear feature, we fit these data to quadratic functions. We gain tiny coefficients of the quadratic terms for both pure InAs/GaAs and alloyed InGaAs/GaAs QDs ($\xi_{3,4}^\text{x} \sim 10^{-7} - 10^{-8}$ ns/MPa$^2$, see fitted data in Fig. \ref{fig-fig6}), which illustrate both intrinsic lifetimes and their asymmetries are linear against [110] stress. In striking contrast to [110] stress applied along the $[100]$ direction gives rise to different response of the lifetime of $\tau_{\text{3x}}$ and $\tau_\text{4x}$ because it causes direct coupling between two bright exciton states (see mechanism in Fig. \ref{fig-fig2}). From Fig. \ref{fig-fig6} we see that both $\tau_{\text{3x}}$ and $\tau_\text{4x}$ exhibit quadratic relations against the applied [100] stress $F$.  As expected, the corresponding lifetime asymmetry $\delta\tau_\text{x}$ also displays a quadratic function of $F$. The similar features can also be found for transition from biexciton to exciton states.

The extrinsic lifetime asymmetries have more complicated response behaviours to applied stress and will exhibit strong nonlinearity even under weak force due to direct coupling between the two bright states
induced by the $C_1$ symmetry potential $V_1$. Fig. \ref{fig-fig7} shows  atomistic calculated results. In the pure InAs/GaAs QD the $V_1$ potential is absent, and thus $T_\text{3x}$ and $T_\text{4x}$ should possess perfect linear functions of $F$ for stress applied along the $[110]$ direction and quadratic functions for stress applied along the [100] direction. The results shown in Fig. \ref{fig-fig7} well support  the theoretical prediction. These stress-responses are identical to $\tau_\eta$, as shown in Fig. \ref{fig-fig6}b,d, in the sense that $\theta = 0$. However, for alloyed InGaAs/GaAs QD, strong nonlinearity behaviours of both mean extrinsic lifetimes and extrinsic lifetime asymmetries are expected for stress applied along [100] and [110] directions. Since $T_i(F) = \tau_i(F)$ (see Eq. \ref{eq-deltaTF}), we can observe the strong nonlinearity in both the extrinsic lifetimes $T_{i\eta}$ and the extrinsic lifetime asymmetries $\delta T_\eta$. Moreover, while the extrinsic lifetime asymmetries are generally very small due to wave function mixing effect, we find that the extrinsic lifetime asymmetries in Fig. \ref{fig-fig7} c-d, and the intrinsic lifetime asymmetries in Fig. \ref{fig-fig6}c-d, can be pronouncedly enhanced by external stress from tens of ps to 0.2 - 0.7 ns, that is, by at least one order of magnitude. This enhanced asymmetries may lead to direct measurement of lifetime asymmetries with only time-resolved photoluminescnce spectrum, in which the time-resolved spectrum should be fitted using two exponential decay functions. The similar strong nonlinearity effect has also been observed for transition from biexciton to exciton states in a reason discussed above. These results can be well described by the data in Fig. \ref{fig-fig6} and are in fully accordance with our theoretical predictions.

{\it Summary}. We introduced lifetime asymmetry, which is a new concept,  into self-assembled QDs. We revealed that intrinsic lifetimes are fundamental quantities of QDs, which determine the bound of the extrinsic lifetime asymmetries, polarization angles, FSSs, and their evolution under uniaxial external forces. These predictions can be direct measured or extracted from experiments using the state-of-the-art techniques, such as the measured  linear polarization as well as the optical properties of QDs under external forces. We verified these predictions using atomistic simulations. We found that the intrinsic lifetime asymmetries can be of the order of few hundred picoseconds in pure InAs/GaAs QDs, but the extrinsic lifetime asymmetries can be much smaller in alloyed InGaAs/GaAs QDs. However, the lifetime asymmetries are susceptible to external forces and its directions, thus can be more conclusively verified by investigating their behaviours under external forces. These exact relations represent a complete description of the optical properties of QDs. Our findings provide an important basis to deeply understanding properties of QDs.

{\it Acknowledgement}. M.G. is supported by the National Youth Thousand Talents Program (No. KJ2030000001), the USTC start-up funding (No. KY2030000053),
the national natural science foundation (NSFC) under grant No. GG2470000101). X. X. is supported by National Basic Research Program of China under No. 2014CB921003, 
the NSFC under Nos. 11721404, 91436101 and 61675228; the Strategic Priority Research Program of the Chinese Academy of Sciences under No. XDB07030200 and XDPB0803, and the CAS 
Interdisciplinary Innovation Team. J. L. is supported by NSFC under Nos. 61121491, 11474273, 11104264 and U1530401, and National Young 1000 Talents Plan. X. W. is supported 
by NSFC under No. 61404015 and fundamental research funds for the central universities (2015CDJXY300001).

\end{document}